\documentclass[letterpaper,10pt]{article}
\usepackage[letterpaper, margin=1in]{geometry}

\usepackage{epsfig,hyperref,listings,graphicx,multicol,fancyvrb}

\usepackage{xcolor,mdframed}
\usepackage[labelfont=bf]{caption}

\begin{document}

\graphicspath{ {./} }

\date{}

\title{ Constant-Time Wasmtime, for Real This Time\\
\large End-to-End Verified Constant-Time Programming\\ for the Web and Beyond}

\author{
{\rm Garrett Gu}\\
UT Austin
\and 
{\rm Hovav Shacham}\\
UT Austin
}
\maketitle

\thispagestyle{empty}

\subsection*{Abstract}

We claim that existing techniques and tools for generating and verifying constant-time code are incomplete, since they rely on assumptions that compiler optimization passes do not break constant-timeness or that certain operations execute in constant time on the hardware. We present the first end-to-end constant-time-aware compilation process that preserves constant-time semantics at every step from a high-level language down to microarchitectural guarantees, provided by the forthcoming ARM \texttt{PSTATE.DIT} feature. \cite{arm_dit}

First, we present a new compiler-verifier suite based on the JIT-style runtime \texttt{Wasmtime}, modified to compile \texttt{ct-wasm} \cite{watt2019ct}, a preexisting type-safe constant-time extension of WebAssembly, into ARM machine code while maintaining the constant-time property throughout all optimization passes. The resulting machine code is then fed into an automated verifier that requires no human intervention and uses static dataflow analysis in Ghidra to check the constant-timeness of the output. Our verifier leverages characteristics unique to ct-wasm-generated code in order to speed up verification while preserving both soundness and wide applicability. We also consider the resistance of our compilation and verification against speculative timing leakages such as Spectre. Finally, in order to expose \texttt{ct-Wasmtime} at a high level, we present a port of \texttt{FaCT} \cite{cauligi2017fact}, a preexisting constant-time-aware DSL, to target \texttt{ct-wasm}. 

\section{Introduction}
\label{sec:intro}

The constant-time property has always been an important requirement for any reasonably secure system. With the explosion of cryptography and online systems in recent decades, and Brumley and Boneh proving the practicality of timing attacks on real world systems, the need for constant-time cryptographic primitives was introduced. \cite{brumley2005remote} Several cryptographic libraries such as BearSSL \cite{bearssl} and OpenSSL \cite{openssl} found that oftentimes large-scope rewrites were necessary in order to protect against the threat of timing side-channel attacks. 

The existence of Single-Origin Policy and embeddable iframes in modern browsers brought some interesting secrecy challenges, with the need for embedded iframes and the websites surrounding them to be isolated from each other, while still allowing for some amount of interaction. Andrysco, et al. showed that SVG filters in several modern browsers used floating-point numbers, which exhibited strongly data-dependent timing when the value is subnormal, triggering a slow path in the FPU. \cite{andrysco2015subnormal} Mitigating subnormal floating point side channels has been done by flushing all subnormal values to zero, reducing timer precision, or abandoning floating-point calculation altogether. \cite{kohlbrenner2016trusted,kohlbrenner2017effectiveness}

However, note that the act of writing, compiling, and verifying constant-time code has always been a guess-and-check activity. Programmers "know" that their program is constant-time when it avoids operations already known to be unsafe, or when they measure its runtime against various values and it seems to always take the same amount of time, but this "knowledge" is almost never backed by any guarantees beyond guesses of hardware behavior. Our \texttt{ct-wasm} compiler suite exposes new hardware guarantees on constant-time behavior in a practical and usable way, and as a result leaves very little room for doubt on whether the resulting code truly exhibits constant-time behavior.

Our approach considers carefully the transformations being applied to our code throughout the compilation process, then statically analyzes the final machine code to ensure it maintains the same constant-time guarantees as the original code. Our approach differs from previous approaches in that it only relies on guarantees given by the hardware provider --- we do not assume constant-time preservation of any compiler optimizations, nor do we take for granted that certain "trivial" operations are constant-time without being told so by the architecture (such as integer addition). 

\section{Background}
\subsection{The Problem with Current Constant-Time Verification Techniques}
\label{sec:theproblem}
Let's consider \texttt{libfixedtimefixedpoint}, a library developed by Andrysco et al., mitigating floating-point timing side channels by providing a fixed-point alternative. \cite{andrysco2015subnormal} In 2016, Almeida et al. presented \texttt{ct-verif}, which verified the constant-timeness of \texttt{libftfp}. \cite{almeida2016verifying}

\begin{figure*}
  \includegraphics[width=\textwidth,height=7cm]{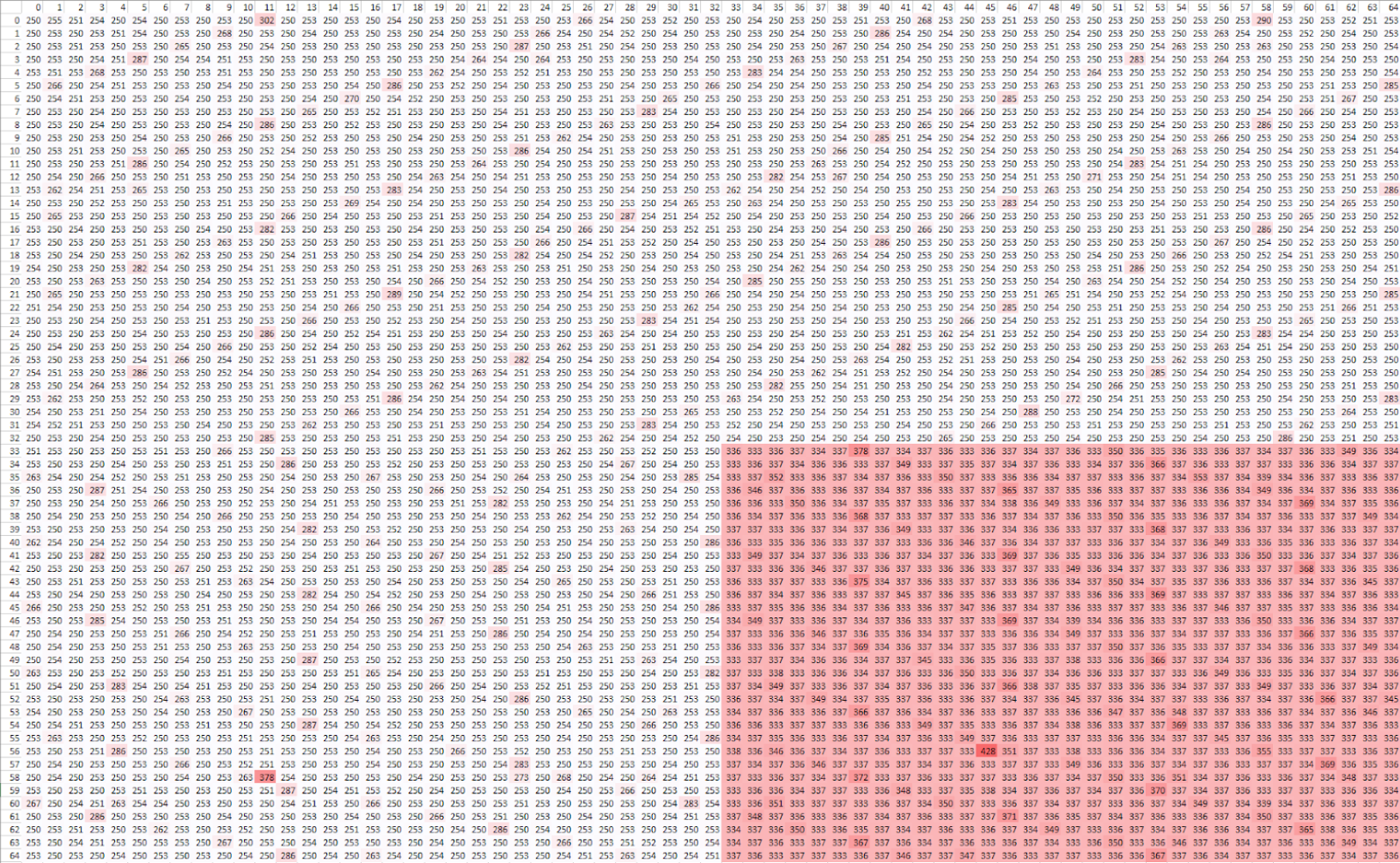}
  \caption{A generated matrix of the amount of time in microseconds it takes to multiply 10,000 m-wide by n-wide integers on the Raspberry Pi 3 B. The number of bits set in the left-operand increasing from left to right, and the number of bits set in the right-operand increasing from top to bottom. We clearly see a difference in timing when both operands are over 32 bits wide. We also note that some diagonal patterns are likely caused by preemption and are not indicative of timing variation. }  \label{fig:mtx}
\end{figure*}

However, when compiled to ARM machine code and run on a Cortex-A53 or A55 core (used by several modern smartphone processors as their LITTLE cores), the software clearly exhibits wild timing variability, as evidenced by a testbench program, \texttt{dudect}. (\autoref{fig:dudect}) The issue with the library (and by extension, the verification efforts applied to it) was that it applied verification at the compiler IR level, and as a result made assumptions about the timing characteristics of the lowered program on the underlying architecture --- assumptions that may have held on the hardware it was intended for and tested on (2015 Intel and AMD processors), but clearly do not hold on other architectures. In this particular instance, the assumption was that 64-bit integer multiplication was constant-time, which is in fact not the case on the Cortex-A53 core; the core expends one extra cycle when both operands happen to be wider than 64 bits. (See \autoref{fig:mtx}) While it is fairly common to see this type of behavior on Thumb or embedded chips \cite{pornin_2018}, we were unable to find any literature on this integer multiplication timing variability on modern AArch64 chips, and ARM's own Cortex-A53 Manual does not mention the behavior. \cite{arm_a53} We were also able to reproduce the same behavior on half the cores on a modern Android smartphone equipped with a Qualcomm Snapdragon 855 CPU. 

\begin{figure}
\begin{mdframed}[linecolor=gray]
\begin{Verbatim}[fontsize=\footnotesize]
meas:  0.99 M, max t:   +1.79, max tau: 1.79e-03, (5/tau)^2: 7.78e+06. For the moment, maybe constant time.
meas:  1.81 M, max t: +454.21, max tau: 3.38e-01, (5/tau)^2: 2.19e+02. Probably not contant time.
meas:  2.71 M, max t: +773.34, max tau: 4.70e-01, (5/tau)^2: 1.13e+02. Definitely not constant time.
\end{Verbatim}
\caption{The ouput gathered by \texttt{dudect} when running on \texttt{fix\_sin} from LibFTFP. The two sample groups being compared are zero and random. \texttt{dudect} almost instantly concludes the function is not constant time. }
\label{fig:dudect}
\end{mdframed}
\end{figure}

What's also worrying is that the library is technically not even guaranteed to execute in constant-time on the hardware it was designed for and tested on; for instance, Intel and AMD never guaranteed that the \texttt{imul} instruction doesn't have some subtle edge case that causes it to occasionally take a couple additional cycles to execute depending on input. In this particular instance, we may never disprove the existence of such an edge case, since the input space of 64-bit \texttt{imul} spans $2^{128}$ possibilities. 

In conclusion, for any constant-time verification to hold any weight, it is necessary for the verification to extend all the way down to the bare metal on which the verified code runs. ARM's \texttt{PSTATE.DIT} feature allows us to take advantage of explicit hardware guarantees on timing rather than make guesses on which operations \textit{may be} constant-time.

\begin{figure*}
\begin{center}
  \includegraphics[width=15cm]{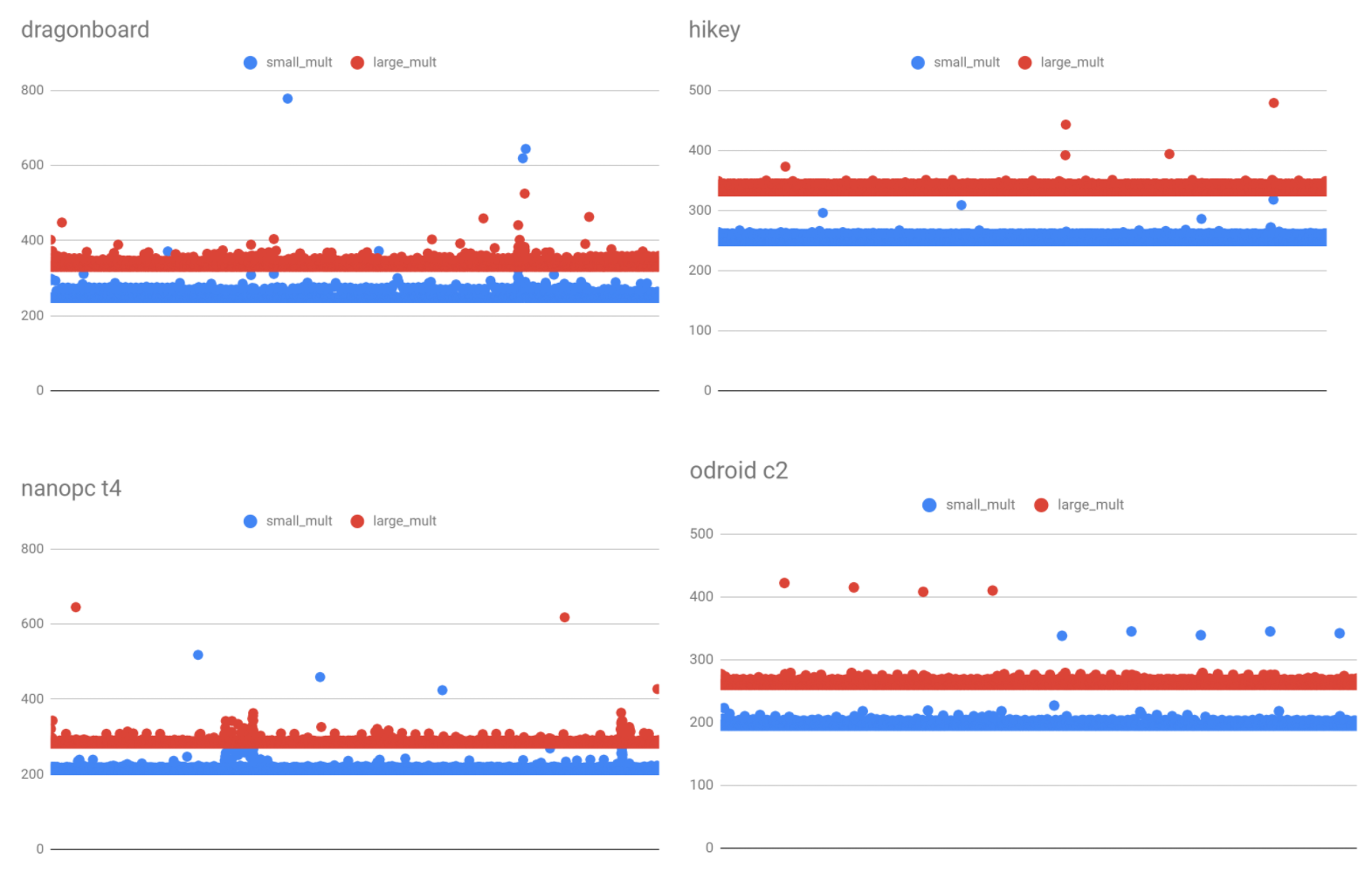}
\end{center}
  \caption{Measured 32-wide vs 64-wide multiplication on Cortex A-53 devices. The vertical axis is the number of microseconds measured by Linux's gettimeofday() syscall which it takes these devices to multiply 10,000 integers, and the horizontal axis denotes the nth sample taken. }
  \label{fig:a53}
\end{figure*}


\subsection{Straight-Line Constant-Timeness}
In order to achieve data-independent timing, we require our programs to fulfill the following criteria: 
\begin{enumerate}
    \item No secret-dependent branching. Varying the program control flow leads to information leakage through the branch predictor as well as leakage when the two branches are not perfectly identical in timing behavior and side effects (such as cache accesses). 
    \item No secret-dependent memory accesses. These loads/stores may exhibit timing variability depending on the cache or paging state of the system, and may also contribute to future cache timings. 
    \item No secret-dependent instruction timing. For example, floating-point instructions typically demonstrate wild timing variability, and were historically used to break browser isolation guarantees. \texttt{PSTATE.DIT} primarily concerns itself with this criterion.
\end{enumerate}

This is essentially the same leakage model adopted by most previous approaches. \cite{almeida2016verifying,barthe2019formal} Absent some speculative leaks (which we address later in \autoref{sec:spectre}), we claim that a program that provides the above guarantees will not leak secret information through cache effects or execution time. Unlike previous verified constant-time programs, we are able to be confident that our timing assumptions will hold on any future CPUs as well provided they support \texttt{DIT}.

\subsection{What is and isn't covered under DIT}
\label{sec:ditdive}
When the DIT flag is turned off, ARM makes no statement about the timing of any instructions. When the DIT flag is set on compatible processors, ARM requires that: \cite{arm_dit}
\begin{enumerate}
    \item The timing of every load and store instruction is insensitive to the value of the data being loaded or stored.
    \item For certain data processing instructions, the instruction takes a time which is independent of:
    \begin{enumerate}
        \item The values of the data supplied in any of its registers.
        \item The values of the NZCV flags.
    \end{enumerate}
    \item For certain data processing instructions, the response of the instruction to asynchronous exceptions does not vary based on:
    \begin{enumerate}
        \item The values of the data supplied in any of its registers.
        \item The values of the NZCV flags.
    \end{enumerate}
\end{enumerate}

In the manual \cite{arm_dit}, the above requirements are accompanied with a list of cryptographic, general-purpose, and SIMD instructions for which points 2 and 3 apply. Some omissions from the list were seen in earlier attacks and mitigations, for example:
\begin{enumerate}
    \item Floating-point arithmetic instructions are omitted entirely, probably due to the prohibitive performance cost of stretching our every floating-point arithmetic operation to accommodate for worst-case values such as subnormals. However, we have noted that "constant-time" implementations such as Firefox's SVG filter engine assume that floating-point operations exhibit data-independent timing when subnormals are flushed to zero.
    \item Loads/stores are not timing-independent of the addresses being accessed. Although some cryptographic implementations may benefit from using a lookup table, we can safely assume that forcing every memory access to take the same amount of time as a full cache miss may be prohibitively expensive. 
\end{enumerate}

\texttt{DIT} weirdly does not make the following guarantees:
\begin{enumerate}
    \item The architecture specifically requires instructions to exhibit timing independent of \textit{their own} registers, but not any registers which are not explicitly used by the instruction. For instance, an implementation of \texttt{add x0, x0, x1} could depend on the value of \texttt{x2} and still be \texttt{DIT}-compliant, despite potentially leaking secret information through timing. One could argue that this is obvious and could be assumed. However, \texttt{DIT} being a formalization of previously-assumed timing behavior, the supposed obviousness of this timing behavior makes its omission even more puzzling. 
    \item Loads and stores are not explicitly specified to exhibit timing independent of the NZCV flags. \cite{arm_dit} In a real-world program where non-secret-dependent loads and stores may be interspersed with secret-processing instructions, the implication is that special care must be taken to ensure that the NZCV flags do not contain secret-dependent values while the load/store instruction is executed. This is difficult for any programmer to ensure without direct access to the toolchain and in a language where condition flags are abstracted away (such as C). 
    \item ADR and ADRP, both of which simply calculate PC-relative addresses, are not included under \texttt{DIT}. \cite{arm_dit} Since these instructions are often needed to perform quick loads and stores from global variables (which are in fact protected by \texttt{DIT}), we found it a bit odd that the manual provides no timing guarantees for them. 
    \item Branching instructions are (mostly) absent. We would have preferred there to be some guarantee that, for instance, unconditional branches do not depend on the NZCV flags or that conditional branches do not depend on some completely irrelevant register (see point 1 above). However, this is risky since reasoning about branch prediction to mitigate timing side channel attacks such as Spectre is significantly more nuanced. \cite{cauligi2020constant} However, since RET in ARM, which is included under \texttt{DIT}, is semantically just an indirect branch, it is possible to implement all manner of branching using just the RET instruction. It is unclear whether ARM would like developers to replace all their indirect branching with RET instructions with some mechanism reminiscent of Retpoline. \cite{kadir2019retpoline}
\end{enumerate}

In the interest of keeping compilation practical, we assume points 1 and 2. However, we note our approach can still work without assuming point 2 with a very mild complication of the compiler's lowering step. 

\subsection{Availability of DIT}
\label{sec:availability}
The \texttt{DIT} feature is very new; to our knowledge no commercially available ARM CPU offers this feature yet. \cite{arm_cortex} After some experimentation, we found that Apple's new M1 ARM chips either do not support the feature or do not expose it to user-mode processes. \cite{m1_specs} As a result, we cannot expect our verified programs to exhibit data-independent timing on existing chips, only future chips once the feature is available and enabled.

\section{ct-wasm}
In order to enable \texttt{DIT} as a compilation target suitable for cryptographic applications, we began by porting \texttt{ct-wasm}, an extension of WebAssembly with constant-time semantics, to the Cranelift code generator. Then, we ensure that Cranelift converts \texttt{ct-wasm} into \texttt{DIT}-aware constant-time code using verification. 

\subsection{Threat Model}
We assume an adversary who is able to execute arbitrary code on the target machine within the same process as the victim program, can look into microarchitectural side channels such as caches, and has access to precise timers.


\subsection{Background}
\subsubsection{\texttt{ct-wasm} Background}
\texttt{ct-wasm} uses extensions to the WebAssembly type system to verify in linear-time that the input program exhibits constant-time semantics. \cite{watt2019ct} It does so by introducing two types containing secret integer values, \texttt{s32} and \texttt{s64}, and carefully monitoring the information flow to and from variables of these types. For instance, \texttt{untrusted} functions (functions that are verified to have constant-time semantics) are not allowed to convert an \texttt{s64} to an \texttt{i64}. In addition, \texttt{s32} variables are not allowed to be used as memory indices or in floating-point operations, and entire linear memories are defined to either contain either only secret values or only public values. (See \autoref{fig:ctwasmsnippet})

\begin{figure*}
  \begin{verbatim}
  (func $update untrusted (param $inputlen i32) (param $test s32)
    (local $i i32)

    (set_local $i (i32.const 0))
    (block
      (loop
        (br_if 1 (i32.ge_u (get_local $i) (get_local $inputlen)))
          (s32.store8
            (i32.add (get_global $data) (get_global $datalen))
            (s32.load8_u (i32.add (get_global $input) (get_local $i)))
          )
          (set_global $datalen (i32.add (get_global $datalen) (i32.const 1)))
          (if (i32.eq (get_global $datalen) (i32.const 64))
            (then
              (call $transform)
              (set_global $bitlen (i64.add (get_global $bitlen) (i64.const 512)))
              (set_global $datalen (i32.const 0))
            )
          )
          (set_local $i (i32.add (get_local $i) (i32.const 1)))
          (br 0)
      )
    )
  )\end{verbatim}
  \caption{An example of \texttt{ct-wasm} code, in this instance the update subroutine of a SHA-256 implementation. Note that values are defined as either of a public type (e.g. \texttt{i32}) or a secret type (e.g. \texttt{s32}), and arguments to \texttt{load}, \texttt{store}, and branching operations are of type \texttt{i32}. In addition, values loaded from memory must be of a secret type since the memory was defined as secret.}
  \label{fig:ctwasmsnippet}
\end{figure*}

As a result, given a \texttt{ct-wasm} program, it is easy to check which functions "should be" constant-time according to \texttt{ct-wasm} semantics (since they are marked with the \texttt{untrusted} keyword) \cite{watt2019ct}; our goal is to verify that these functions maintain their constant-time semantics after undergoing the Cranelift compilation process, modulo \texttt{PSTATE.DIT}. 

\subsubsection{Cranelift Background}
Cranelift describes itself as a "low-level retargetable code generator," and is most widely used as one of the JIT compilers used by Firefox. \cite{cranelift} We choose to build our \texttt{ct-wasm} compiler on Cranelift (specifically the Wasmtime runtime that uses Cranelift behind-the-scenes) because of two reasons:
\begin{enumerate}
    \item Wasmtime supports ahead-of-time (AOT) compilation. This allows us not only to examine the resulting code at its final stage of compilation, but also gives us control over distributing builds of timing-sensitive software that we have verified to be constant-time. 
    \item Not only does Cranelift feature in Firefox, one of the most popular web browsers, Wasmtime allows it to be embedded within other applications easily as an SFI solution. This allows us to bring end-to-end verified constant-time software to both the web and to cryptographic libraries used by other applications.
\end{enumerate}

Cranelift is developed entirely in Rust; as a result, adding constant-time primitives to the compiler was fairly straightforward, since the Rust compiler is able to give pointers to which sections of the compiler need to be worked on next (for instance, through errors caused by unhandled \texttt{match} cases).

\subsection{Changes to \texttt{ct-wasm}}
Since \texttt{ct-wasm} was released, there have been a few developments in the WebAssembly space that conflicted with the original specification. For instance, a conditional \texttt{select} operation has been added, conflicting with the \texttt{ct-wasm} version of \texttt{select}. We added an \texttt{sselect} operation to \texttt{ct-wasm}, which accepts a \texttt{s32} as its condition rather than a \texttt{i32}, and only selects secret values in order to prevent secrecy leaks. In addition, some bits in the bytecode specification had to be shifted around (such as the memory \texttt{shared} bit conflicting with our \texttt{secret}), so we also modified the bytecode specification; as a result, we feed all our test program inputs to the compiler in the WebAssembly text format rather than the binary format. 

\subsection{Changes to Cranelift}
\label{sec:craneliftchanges}
One of the main changes we made to Cranelift was introducing \texttt{DIT} instructions to the Cranelift IR. For instance, we added an \texttt{iaddDIT} instruction as a semantically-equivalent counterpart to the \texttt{iadd} instruction. 

A DIT instruction is handled differently from a regular instruction in the following ways:
\begin{enumerate}
    \item When \texttt{ct-wasm} is lowered to Cranelift IR initially, we ensure that every \texttt{ct-wasm} instruction that handles values of a secret type (such as \texttt{s32.add}) is translated to a \texttt{DIT} instruction.
    \item During each compilation pass, we ensure that a \texttt{DIT} instruction is never replaced with a non-\texttt{DIT} instruction. 
    \item When optimized Cranelift IR is again lowered to AArch64 machine code, we ensure that every \texttt{DIT} IR instruction is translated to a correponding \texttt{DIT} ARM instruction.
\end{enumerate}

\subsection{Constant-Time in SSA}
\label{sec:compileropt}
Since both Cranelift and Ghidra (which we use to verify our code) process code using SSA, we would like to describe what exactly it means for an SSA program to be considered constant-time. Since SSA is particularly amenable to dataflow analysis, we define our constant-time requirements in terms of a dataflow policy.

\subsubsection{SSA Background}
Cranelift, like most modern compilers, uses SSA form for its Intermediate Representation (IR), since SSA form lends itself easily to dataflow analysis and enables some advanced optimizations. \cite{wimmer2010linear, hack2006register}

An IR in SSA form has the property that every variable is assigned only once, and every usage of a variable is dominated by its assignment in the control-flow graph (CFG). 

Typically the IR is split into \texttt{basic blocks}, stretches of code with no inward branches except to the beginning and no outward branches except from the end. In order to handle values that depend on control flow, the SSA form introduces a \textit{$\phi$ function}, a node that chooses one of its arguments dependent on the block executed prior to entering the current block. 

Rather than explicit $\phi$ nodes, Cranelift instead allows basic blocks to receive \textit{parameters}, which are "passed" by jump instructions entering the block. However, it is trivial to convert an SSA program that uses parameters to one that uses $\phi$ nodes, since we can simply insert a new $\phi$ node assignment at the start of the block for each parameter, assigning the variables being passed in through each jump pointing into the block. (\autoref{sec:ssaexample})

\subsubsection{Converting SSA to DFG}
\label{sec:dfgrednode}
When reasoning about compiler optimizations, we found it helpful to convert each hypothetical SSA function into a Dataflow Graph (DFG), a directed graph expressing the explicit dataflows through the variables in the function. \cite{johnson1993dependence} Cranelift typically also operates on an equivalent DFG when performing optimizations. 

Our DFG formulation, which is derived from the internal DFG representation used by Cranelift, creates a node for roughly each statement in the IR, and colors each node "red" if it can possibly lead to leakage, and "white" otherwise. At the "top" of the DFG we place all values flowing into the function, which can be parameters, accessed global variables, values loaded from memories, or return values from any called functions. 

In this DFG formulation, we define a function as constant-time if there is no path from a secret input into a red node. Therefore, we consider a compilation pass safe if, assuming the original code was constant-time, it cannot possibly introduce a new path from a secret input into a red node.

Our DFG does not include information on the control-flow of the function other than its effects on $\phi$ nodes; this information is not needed since secret-based implicit flows (flows due to the control flow) depend on secret-dependent control flow, which is prevented by our complete ban on secret-dependent branching. 

We consider the following types of nodes in our analysis:
\begin{enumerate}
  \item A load is represented as a red node, with a single inbound connection representing the address, and an outbound connection representing the value being loaded.
  \item A store is represented as two nodes with no outbound connections. One represents the store address, which is always red with a single inbound connection representing the address. The other node accepts the stored value, which is red iff the store is targeting a public memory.
  \item A non-DIT IR instruction is represented as a red node, with its inbound connections representing the instruction's inputs and an outbound connection for the value that was assigned by it. 
  \item A DIT IR instruction is represented as a white node, with inbound and outbound connections defined identically as above.
  \item A function call is represented by one node for each parameter, which is colored red for public parameters and white for secret parameters. In addition, if the output value is used, it is represented as an input into the function DFG.
  \item A $\phi$ node is represented as a white node, with inbound connections representing the possible variables to choose from, and an outbound connection representing the assigned variable. 
\end{enumerate}

\subsubsection{SSA Example}
\label{sec:ssaexample}
Consider the following function expressed in Cranelift IR:
\begin{verbatim}
function u0:1(i64 vmctx, i64, i32, i32, i32) -> i32 system_v {
  block0(v0: i64, v1: i64, v2: i32, v3: i32, v4: i32):
      brnz v2, block3
      jump block4

  block4:
      v11 = iconst.i32 42
      v12 = iaddDIT v11, v3
      jump block2(v12)

  block3:
      v13 = iconst.i32 1374
      v18 -> v13
      v14 = imul.i32 v4, v13
      jump block2(v14)

  block2(v10: i32):
      v15 = iconst.i32 4
      v16 = call fn0(v0, v0, v10, v15)
      v5 -> v16
      jump block1

  block1:
      return v5
}
\end{verbatim}

We first rewrite \texttt{block2} to use a $\phi$ instructions instead of a parameter:
\begin{verbatim}
block2:
  v10 = phi(v12, v14)
  v15 = iconst.i32 4
  v16 = call fn0(v0, v0, v10, v15)
  v5 -> v16
  jump block1
\end{verbatim}

The corresponding DFG for the function is shown below (assuming the function's return value is public and the \texttt{fn0}'s parameters and return value are all secret):

\includegraphics[clip,trim=2cm 6cm 8cm 5cm,width=15cm]{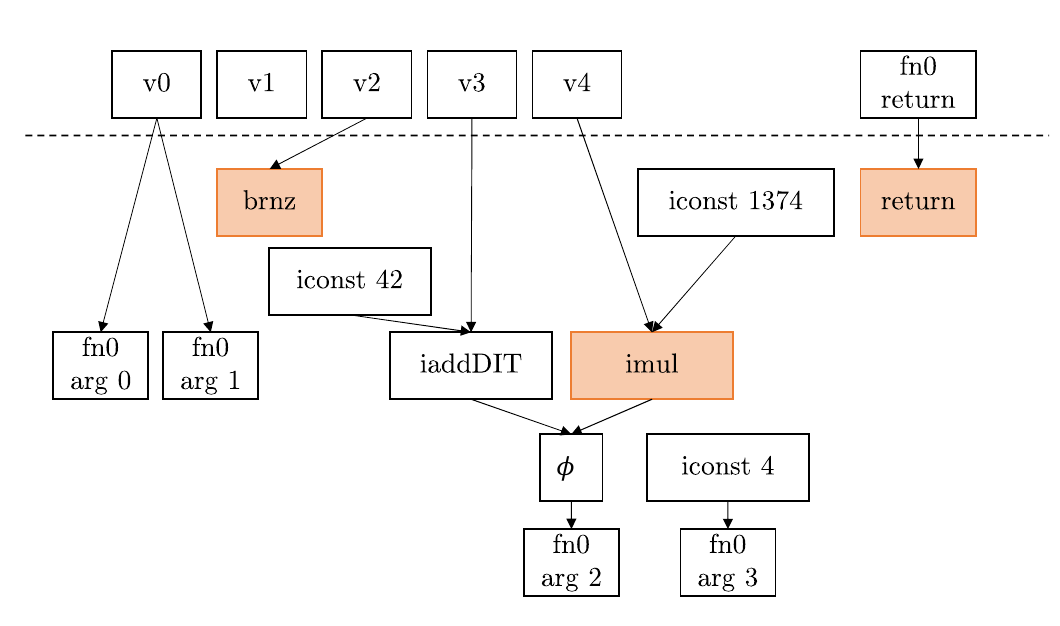}

Note that in the above DFG, \texttt{v2} and \texttt{v4} must be public since they both have a path to a red node, but \texttt{v3} may be secret without breaking constant-timeness.

Using this DFG formulation, we can go through some optimization passes of Cranelift and showing that they are constant-time-safe. Note that this is not a complete list of all the optimizations done by Cranelift, since the list is constantly evolving and already quite long. In addition, we do not rely on any of this reasoning for our verification to be sound, since we treat Cranelift as a black box. 

\subsubsection{DCE}
We claim that the DCE (Dead Code Elimination) pass in Cranelift preserves constant-time semantics. Since we have limited constant-time evaluation to only properties of the DFG, it is very easy to show that DCE cannot introduce timing violations. This is because Cranelift's DCE function only leads to nodes being removed from the DFG, never introduced; therefore, if the original DFG did not contain a path from a secret input to a red node, then the optimized DFG will not either.

\subsubsection{LICM}
We claim that the Loop-Invariant Code Motion (LICM) pass in Cranelift preserves constant-time semantics. Essentially, Cranelift performs LICM by finding instructions within a loop that are invariant with respect to the loop body, then moving them above the loop entry. (\autoref{fig:licm})

\begin{figure}
\begin{mdframed}[linecolor=gray]
\begin{multicols}{2}
\begin{verbatim}
function u0:1(i64 vmctx, i64, i32) -> i32 {
    block0(v0: i64, v1: i64, v2: i32):
        v7 -> v2
        v4 = iconst.i32 0
        v5 = iconst.i32 0
        jump block3(v5, v4)

    block3(v6: i32, v13: i32):
        v8 = icmp uge v6, v7
        v9 = bint.i32 v8
        brnz v9, block2
        jump block5

    block5:
        v10 = iconst.i32 42
        v11 = iconst.i32 1
        v12 = iadd.i32 v6, v11
        jump block3(v12, v10)

    block2:
        return v13
}
\end{verbatim}
\begin{verbatim}
function u0:1(i64 vmctx, i64, i32) -> i32 {
  block0(v0: i64, v1: i64, v2: i32):
      v7 -> v2
      v4 = iconst.i32 0
      v5 = iconst.i32 0
      v10 = iconst.i32 42
      v11 = iconst.i32 1
      jump block3(v5, v4)

  block3(v6: i32, v13: i32):
      v8 = icmp uge v6, v7
      v9 = bint.i32 v8
      brnz v9, block2
      jump block5

  block5:
      v12 = iadd.i32 v6, v11
      jump block3(v12, v10)

  block2:
      return v13
}
\end{verbatim}
\end{multicols}
\caption{An example of LICM being applied to a program. Node that \texttt{v10} and \texttt{v11} were both moved outside of the loop body, yielding a performance increase.}
\label{fig:licm}
\end{mdframed}
\end{figure}

However, note that rearranging the control flow in this way does not affect the equivalent DFG of the function, since we only care about the connections between nodes, not their ordering. Therefore, LICM is trivially constant-time preserving, since we define constant-time with respect to the DFG, not the CFG.

\subsubsection{Peephole Optimization}
As currently implemented, peephole optimization preserves constant-time semantics in our modified version of Cranelift. However, this is a fairly vacuous statement, since we simply do not allow peephole optimizations to operate on DIT instructions at all. For example, consider the following optimization, expressed in Cranelift's DSL for peephole substutition:
\\
\begin{verbatim}
(=> (when (iadd $x $C)
  (fits-in-native-word $C))
  (iadd_imm $C $x))
\end{verbatim}

The above optimization is combined with another peephole optimization to perform constant-folding, since \texttt{\_imm} instructions can trivially be folded into each other. Since \texttt{iadd} is a red node because it is not explicitly DIT, all inputs into the node must already be provably non-secret dependent in order for the original program to be constant-time. Therefore, since the new subgraph does not lead to "leakage" where the original subgraph does not, this peephole optimization is safe.

However, the following hypothetical peephole optimization is not safe:
\begin{verbatim}
(=> (when (iaddDIT $x $C)
  (fits-in-native-word $C))
  (iadd_imm $C $x))
\end{verbatim}

This is because it is replacing a white node with a red node -- the non-constant input to iaddDIT is now being fed into a red node, \texttt{iadd\_imm}, where it was not previously. 

We consider it future work to create DIT versions of these peephole optimizations and prove their correctness.

\subsubsection{GVN}
Cranelift's Global Value Numbering (GVN) pass is currently extremely simple, only implementing a simple redundant subexpression elimination pass. Essentially, it uses value numbering to find variables that provably represent equivalent subexpressions, and replaces the one appearing later in the control-flow with an alias. (\autoref{fig:gvn}) From a DFG perspective, it finds nodes whose parent subgraphs are identical in the DFG, and creates an alias from one node to another. Note that, for instance, \texttt{iadd 2, 3} is considered a different subexpression than \texttt{iaddDIT 2, 3} and \texttt{iadd 3, 2} since Cranelift's GVN pass currently does not consider the semantics of instructions, only their opcodes. 

\begin{figure}
  \begin{mdframed}[linecolor=gray]
\begin{multicols}{2}
\begin{verbatim}
function u0:0(i64, i64, i32, i32) {
  blk0(v0: i64, v1: i64, v2: i32, v3: i32):
      v5 = iconst.i32 0
      v6 = iconst.i32 2
      v7 = imul v2, v6
      v8 = iadd v7, v3
      v9 = iconst.i32 2
      v10 = imul v2, v9
      v11 = iadd v10, v3
      v12 = iadd v8, v11
      v4 -> v12
      jump block1

  blk1:
      return v4
}
\end{verbatim}
\begin{verbatim}
function u0:0(i64, i64, i32, i32) -> i32 {
  blk0(v0: i64, v1: i64, v2: i32, v3: i32):
      v5 = iconst.i32 0
      v6 = iconst.i32 2
      v9 -> v6
      v7 = imul v2, v6
      v10 -> v7
      v8 = iadd v7, v3
      v11 -> v8
      v12 = iadd v8, v8
      v4 -> v12
      jump block1

  blk1:
      return v4
}
\end{verbatim}
\end{multicols}
\caption{An example of GVN being applied to a program. Note that several redundant computations were instead replaced with aliases, which produce zero overhead in the final output.}
\label{fig:gvn}
\end{mdframed}
\end{figure}  

We claim that Cranelift's GVN pass does not break constant-time semantics. Consider a function DFG $F$. Now, let's suppose nodes $a$ and $b$ are proven identical by Cranelift's simple GVN pass, which implies they are connected to the same function inputs through an equivalent set of intermediate nodes. Then, Cranelift will replace $b$'s assignment node with an alias node pointed to by $a$. (It is possible that $b$'s original parent nodes are now eligible for DCE.) Let's call the modified DFG $F'$. 

Now suppose for sake of contradiction that there is a path from one of the input nodes in $F'$ to a red node, but there is no such path in $F$. Note the path must contain the newly created edge from $a$ to $b$ (let's call it $e$), since this is the only new edge added to $F'$. Then we can split the path into two subpaths, one going from the input to $a$, and the other going from $b$ to a red node. Note the second path must also exist in $F$ since the only edge in $F'$ that's not in $F$ points to $b$. Also note that $F$ must also have an equivalent path to the first path since the path obviously cannot contain $e$. Therefore, $F$ contains another path from the input node terminating in $b$ rather than $a$, since we already determined that $a$ and $b$ are connected in $F$ to the same inputs through an equivalent set of intermediate nodes. Therefore, we can combine the two paths to find a path in $F$ from an input to a red node. So we arrive at a contradiction. 

\subsubsection{Other Optimizations}
Cranelift also performs more optimizations, but these are mostly more platform-specific or easier to reason about than the ones we already mentioned. To reiterate, it is not necessary to prove all Cranelift optimizations correct since our verifier treats the entire compiler as a black box. 

\subsection{Verifying Cranelift Output}
\label{sec:verifying}
Wasmtime and Cranelift, combined, are extremely complex pieces of software consisting of around 750,000 lines of constantly-evolving Rust code. As a result, it is infeasible to comb through each line of code on every update to ensure that every compiler optimization preserves CT semantics. 

\begin{figure*}
  \begin{mdframed}[linecolor=gray]
  \begin{verbatim}
{
  "functions":
    [
      {"paramSecrecy":[true],"returnSecrecy":[],"trusted":false},
      {"paramSecrecy":[false],"returnSecrecy":[],"trusted":false}
    ],
    "globalsOffset":64,
    "globalsSecrecy":[],
    "memories":[{"imported":true,"secret":false}],
    "memoriesOffset":28
}\end{verbatim}
  \caption{A sample JSON data structure provided to the verifier. The string is automatically included in the binary by the compiler through a custom section. In this sample module, there are two void functions, one accepting a single secret parameter, the other accepting a single public parameter, and both functions are marked \texttt{untrusted}, meaning they will be checked by the verifier. The module also contains no globals and a single non-secret memory, located at 28 bytes relative to the \texttt{vmctx} struct. }
  \label{fig:ghidrajson}
\end{mdframed}
\end{figure*}

We opted instead to verify the AArch64 machine code generated by the Cranelift code generator. In order to do this verification, we verify each function individually, and specify to the verifier the secrecy traits of every input to each function. (See \autoref{fig:ghidrajson}) Our verifier then constructs the dataflow graph as described earlier and ensures that secrecy taint never spreads into a "red" node. We implement our verifier, JANT (JIT Analyzer for Non-secret-dependent Timing) in about 1,200 lines of Java code using the dataflow analysis features of Ghidra. 

Note that JANT is not concerned with the \textit{functional correctness} of the resulting code, since bugs can easily be found with rigorous testing or, if needed, a formal correctness verifier. JANT verifies that, regardless of whether or not the translation adheres to the functional semantics of the input program, the resulting machine code adheres to the \textit{constant-time semantics} of the input program, meaning functions defined as \texttt{untrusted} will not leak secret values through timing or cache effects.

\subsubsection{Ghidra Background}
We perform constant-time verification of the machine code using Ghidra, a reverse-engineering framework open-sourced by the NSA. \cite{rohleder2019hands} Ghidra is most well-known for its included decompiler, which, given an arbitrary binary, produces equivalent code in a high-level pseudo-C language. Since it does support several different architectures, the first step in its analysis routine is to lift the binary's machine code into \textit{pcode}, a DSL with SSA semantics and a very limited instruction set. \cite{pcode} This also also exposes side effects of instructions explicitly. For instance, an arithmetic operation in machine code seems to only affect its output register, but in reality may set processor flags depending on input. In pcode, such side effects are made explicit, ensuring that the resulting dataflow graph is not missing any connections. 

Using Ghidra also comes with the benefit of portability; JANT does not use any ARM-only constructs besides a simple allowlist of instructions (\texttt{DIT}-instructions) that are able to operate on tainted values. Thus, when other architectures begin providing constant-time guarantees, our approach will easily extend to them as well, since we can simply lift into pcode and apply the same analyses. 

\subsubsection{Intended Usage and Goals of Verification}
The intended use case of JANT is as a bugfinding tool for Cranelift, discovering how timing leakages can be introduced by both current and future optimization passes introduced to Cranelift. In addition, the binaries generated by Cranelift can be run directly without the original WebAssembly code used to generate them, with confidence granted by the verifier that they will not leak secrets through timing side-channels. 

With that in mind, we designed the verifier with the following goals in mind:
\begin{enumerate}
  \item The verifier should \textit{never} admit an unsafe program as safe. (Soundness)
  \item The verifier should \textit{usually} admit a safe program as safe. (Completeness)
\end{enumerate}

Clearly, point 1 is very important since we want to inspire confidence that JANT-verified code will not leak secrets. Point 2 is also important (but slightly less so) since we would like JANT to be useful to as many programs as possible.

As a result of our priorities, we take an extremely conservative approach to verification, trying to eliminate false negatives (unsafe programs marked as safe) with the secondary goal of minimizing false positives (safe programs marked as unsafe). For instance, throughout verification we are extremely careful not to miscategorize a potentially secret value as a public value, but we are relatively less careful about the reverse; this is because the first miscategorization would yield a false negative while the second one would yield only a false positive. 

\subsubsection{Basic Verification Algorithm}
\label{sec:basicverif}
At a high-level, our verifier uses the following algorithm to find timing leakages in a function:
\begin{enumerate}
  \item Find all red nodes in the function. (\autoref{sec:dfgrednode})
  \item For each red node, construct the \textit{generalized expression tree} corresponding to each of its operands. 
  \item For each leaf node in the generalized expression tree, assert that the leaf is not secret-dependent.
\end{enumerate}

\begin{figure}
  \begin{mdframed}[linecolor=gray]
\begin{multicols}{2}
\includegraphics[clip,trim=4.5cm 7.75cm 14.5cm 5.25cm,width=0.40\textwidth]{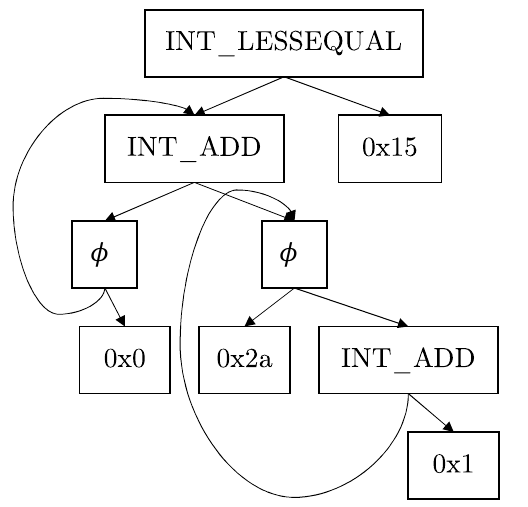}
\begin{verbatim}


  i32 i = 42;
  i32 j = 0;

  while(i < limit) {
    j += i;
    
*  if(j < 21) {
      ...
    }

    i += 1;
  }
\end{verbatim}
\end{multicols}
\caption{An example of a generalized expression tree generated by JANT for the branch condition of the starred if statement on the right. Note that even though there are cycles caused by the $\phi$ nodes, every leaf of the generalized expression tree is constant (and therefore non-secret-dependent); we conclude the branch condition is not secret-dependent.}
\label{fig:ast}
\end{mdframed}
\end{figure}

We define a generalized expression tree as an expression tree that may include cycles (which correspond to Phi nodes in the SSA construction), and we construct them by simply following SSA assignments back until we reach a leaf node, which are either constant or one of the input nodes we place on "top" in \autoref{sec:dfgrednode}. We can think of this as a subgraph of the overall DFG of the function, but with the edge directions reversed. (\autoref{fig:ast})

\pagebreak
\subsubsection{What About Implicit Flows?}
One concern that comes up with this type of relatively simple dataflow analysis is the existence of implicit flows from secret values. For instance, consider the following pseudocode:

\begin{verbatim}
  if(i == 0) {
    v = 42;
  } else {
    v = 21;
  }
\end{verbatim}

Here, the generalized AST constructed for the variable \texttt{v} after the if statement would look something like $\phi(const(42), const(21))$, which loses the implicit dataflow coming from \texttt{i}. However, we can ignore this type of implicit flow due to the properties of constant-time code, which require all branching to be non-secret dependent. Since we separately check that all branching is based on untainted values, we know that all implicit flows must originate from untainted varnodes, therefore we will not misclassify a tainted value as an untainted value as a result of implicit dataflows from program structure.

However, there is one type of secret-dependent conditional branching we do want to allow. Ghidra pcode allows for \textit{pcode relative branching}, which is defined as branching within the same instruction. \cite{pcode} For instance, \texttt{csel} is a conditional select instruction covered under \texttt{DIT}, but its pcode semantics include a secret-dependent conditional branch by design. For example, consider the following (cleaned-up) translation from a \texttt{csel} instruction into Ghidra pcode. This translation is problematic since it contains both an implicit flow from \texttt{ZR} to \texttt{x1} and a secret-dependent semantic conditional branch that is still constant-time according on \texttt{DIT}. (Note that MULTIEQUAL is the name Ghidra assigns to $\phi$ nodes.)

\begin{verbatim}
csel x1,x1,x2,ne
    $U0:1 = BOOL_NEGATE ZR
    $U1:8 = COPY x1
    CBRANCH <0> , $U0:1
    $U1:8* = COPY x2
<0>:
    x1 = MULTIEQUAL $U1:8*, $U1:8
\end{verbatim}

We solve this issue by first searching the program for all pcode \texttt{CBRANCH} instructions, and rejecting the program if a secret-dependent inter-instruction \texttt{CBRANCH} exists, since this would most likely translate to a secret-dependent branch in the source program. In contrast, we allow intra-instruction secret-dependent conditional branches to exist but we preemptively mark all $\phi$-nodes in the same instruction as tainted to avoid missing implicit flows as a result of the pcode-relative branch. 

\subsubsection{Verifying Loads/Stores}
\label{sec:verifloadstore}
One property of \texttt{ct-wasm} code we utilize in JANT is the fact that explicit loads and stores are statically defined to yield either secret or public values. The two types of load/store instructions we consider are accesses to global variables and linear memories. A load from a global variable load could yield a secret value only if it is defined as a \texttt{s32} or \texttt{s64}. Similarly, a linear memory load could yield a secret value only if it is marked \texttt{secret}.

Of course, our main goal is to never misclassify a secret value as a public value. As a result, we also need to make sure that a tainted value is never written to a non-secret global variable or a non-secret memory, since any load from the same location would later be classified as untainted. 

JANT requires loads and stores to fall cleanly into exactly one of the following categories:
\begin{enumerate}
  \item A load or store into a global variable.
  \item A load or store at a 32-bit untainted offset relative to the base address of a defined linear memory.
  \item A load from a constant address. The yielded value is considered public since we do not allow storing into constant addresses, so the loaded value is always baked into the binary itself. 
  \item A load or store into a constant offset relative to the location of the stack pointer at the start of the function. We enforce the constant offset to simplify taint tracking of spilled register values. WebAssembly does not put arrays or buffers on the stack by design.
\end{enumerate}

\begin{figure}
  \begin{mdframed}[linecolor=gray]
\includegraphics[clip,trim=0.5cm 7.5cm 6cm 5cm,width=0.95\textwidth]{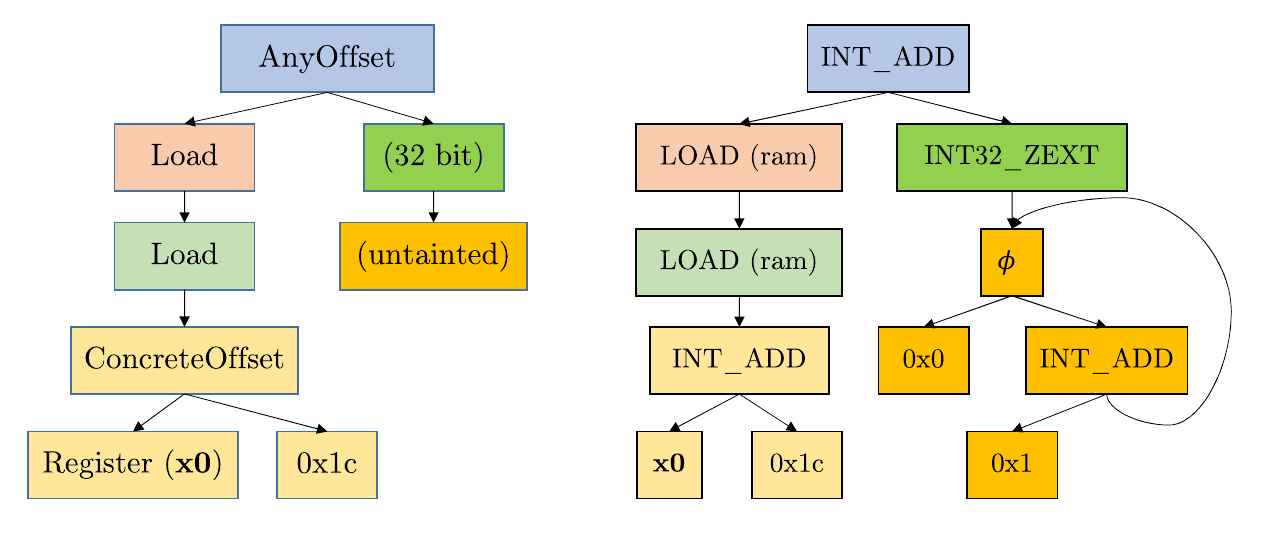}
\caption{An example of a pattern match being performed by JANT. In this case, the address is loaded at an offset relative to the base of an imported linear memory, which in Cranelift translates to a double pointer within the \texttt{vmctx} struct, which is passed into the function through \texttt{x0}. Note that we check that the offset is not secret-dependent using, again, the algorithm described in \autoref{sec:basicverif}.}
\label{fig:patternmatch}
  \end{mdframed}
\end{figure}

We match against these patterns by applying a simple tree pattern matching algorithm against the generalized expression tree of the target address. (\autoref{fig:patternmatch})

\subsubsection{Verifying the Function Call Interface}
\label{sec:veriffuncs}
We also need to verify that secrecy is respected on function interface boundaries. For instance, a secret value could be passed into a function expecting a public value, then the callee could then branch dependent on the secret. 

In order to check that function calls do not leak secrets, we first ensure that every branch is resolved statically so that we do not miss any possible function calls. Due to the very structured control-flow of WebAssembly, the overwhelming majority of branches have only one possible destination. Then, we consider each branch individually. If its destination is within the function currently being analyzed, we do not need to continue checking, since this branch is already considered when constructing the dataflow graph. Otherwise, we check that its destination is the entry point of an untrusted function, then check each one of its public parameters to ensure they are not tainted. 

We also need to handle return values. Ghidra represents return values (and other registers potentially cobbled by the function call) using a special \texttt{INDIRECT} instruction pointing to the pcode instruction causing the cobble. We handle return value secrecy by simply following \texttt{x0} \texttt{INDIRECT} instructions that depend on a function call and checking against the function signature. Finally, we check that the currently analyzed function does not return a tainted value as public. 

\subsection{Verifying Against Spectre}
\label{sec:spectre}
The goal of this paper is not to protect WebAssembly code against Spectre (this has been done by \cite{narayan2021swivel}), and we do not claim to be secure against all variants of Spectre. However, we note that the nature of our conservative verification process naturally prevents several variants of Spectre from being mounted against software verified by JANT. 

In addition, JANT tends not to break with the inclusion of some hardware Spectre mitigations, since semantically transparent instructions such as ARMv8.3's pointer-authentication instructions or ARMv8.5's forthcoming SB instruction produce no corresponding pcode instructions in Ghidra. When Spectre mitigations are added to Cranelift, we can likely benefit with very little changes required to JANT.

\subsubsection{Spectre-PHT}
Spectre-PHT (Bounds Check Bypass) is an attack on the Pattern History Table (PHT), where the attacker causes the CPU to mispredict a branch and leak a secret speculatively. \cite{kocher2019spectre} (See \autoref{fig:spectrepht})

\begin{figure}
  \begin{mdframed}[linecolor=gray]
\begin{verbatim}
    uint32 i = ...; // suppose i is controlled by an attacker
    public uint8[8] buf1;
    secret uint8[8] buf2;
    if(i >= 8) {
      // out of bounds!
      return 0;
    }
    return buf1[i];
\end{verbatim}
\caption{A code snippet vulnerable to Spectre-PHT. In straight-line execution, the program is memory safe. However, in speculative execution, an attacker can mistrain the PHT, bypass the bounds check, and cause the program to speculatively access past the bounds of buf1 to leak a secret value from buf2.}
\label{fig:spectrepht}
\end{mdframed}
\end{figure}

However, we claim that constant-time code verified by JANT is safe against this class of attacks. The only situation in which we allow a non-statically-determined offset in a load or store is in the case of linear memories. (See \autoref{sec:verifloadstore}) However, on 64-bit architectures like AArch64, Wasmtime (and by extension, \texttt{ct-Wasmtime}) keeps linear memories separate by ensuring that memories are separated by at least the size of a 32-bit address space, and that they are only accessed through a 32-bit offset. Therefore, we do not rely on bounds checks anywhere to keep secret and public memories separate. If an attacker were to speculatively read out of bounds, they would simply read a different public value or read into the unmapped guard pages following the public linear memory since their offset is limited to 32 bits. 

\subsubsection{Spectre-BTB}
Spectre-BTB is an attack on the indirect branch predictor, referred to here as the BTB (Branch Target Buffer). Depending on the exact details of how indirect branches are predicted by the CPU, the attacker typically mistrains the BTB in order to cause speculative control flow to redirect to a gadget at an arbitrary address. \cite{kocher2019spectre,zhang2020exploring}

JANT is fairly inflexible when it comes to indirect branches. This is because due to the structured control flow of WebAssembly, the Wasm semantics essentially only allow expressing indirect branches through special \textit{Tables}, which we found are not very useful for writing cryptography code; therefore JANT does not even allow loads/stores out of tables. (\autoref{sec:verifloadstore}) Therefore, for the most part we can avoid indirect branching altogether.

\begin{figure}
  \begin{mdframed}[linecolor=gray]
\centering
\includegraphics[width=7cm]{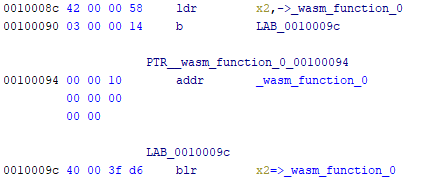}
\caption{A function call generated using a loaded indirect branch by Cranelift, most likely to avoid calculating branch target offsets. During normal execution, the indirect branch has only one possible target. }
\label{fig:cranelift_dumb}
  \end{mdframed}
\end{figure}

Unfortunately, Cranelift's code generation for ARM is still a bit unoptimized at the moment, and it generates indirect branches for function calls unnecessarily. However, the code being generated still has the property that each indirect branch only has one potential destination. (\autoref{fig:cranelift_dumb})

Since JANT requires statically resolving each indirect branch target to a single potential destination in order to verify the function call interface (\autoref{sec:veriffuncs}), we claim it is not possible to mistrain the BTB by executing code that we verified, assuming multiple indirect branch instructions do not alias to the same BTB entry. On modern processors and reasonably-sized sandboxed programs, this is a reasonable assumption to make, but this is an example of where our Spectre avoidance is not completely sound. 

Of course, the attacker may have control over a separate sandbox or process from which they can mistrain the BTB, but this is easily mitigated by simply flushing the BTB on sandbox transitions. \cite{narayan2021swivel}

\subsubsection{Spectre-RSB}
Spectre-RSB is an attack on the RSB, a small stack-like buffer that keeps track of where \texttt{call} instructions were recently executed so that the targets of \texttt{ret} instructions can be predicted accurately most of the time. The attacker uses their own sandbox to overflow or underflow the RSB to speculatively redirect control flow after a \texttt{ret}. \cite{koruyeh2018spectre} Previous Spectre-RSB mitigations on Intel-based systems used shadow stacks, since Intel CET guarantees that shadow stack mismatches will halt speculative execution past a \texttt{ret}. \cite{narayan2021swivel}

The equivalent feature to shadow stacks on ARM is return address authentication, where a cryptographic authentication code is injected into the unused upper-bits of the return address at the beginning of the function, and the upper bits are later verified prior to returning from the function. 

We implemented return address signing to Cranelift with little difficulty, and the inclusion of the feature required zero changes to JANT for the verification to continue working. Although ARM does not provide the same guarantees as Intel that we need to conclusively rule out Spectre-RSB, we assume that return address authentication will most likely prevent misspeculation through the RSB on most chips. In addition, if ARM releases a different official hardware mitigation for RSB, we claim that adding support for said mitigation into Cranelift will likely not require significant changes to JANT.

\subsection{Evaluation of \texttt{ct-Wasmtime}}
We focus our evaluation on the runtime performance cost of using \texttt{ct-Wasmtime} over running the stripped, non-guaranteed-constant-time equivalents of the same programs. We do not consider the binary size cost of using \texttt{ct-wasm}, since this was already done by Watt et. al. \cite{watt2019ct} 

\subsubsection{Runtime Overhead}
\begin{figure}
  \begin{mdframed}[linecolor=gray]

\centering
\includegraphics[width=\columnwidth]{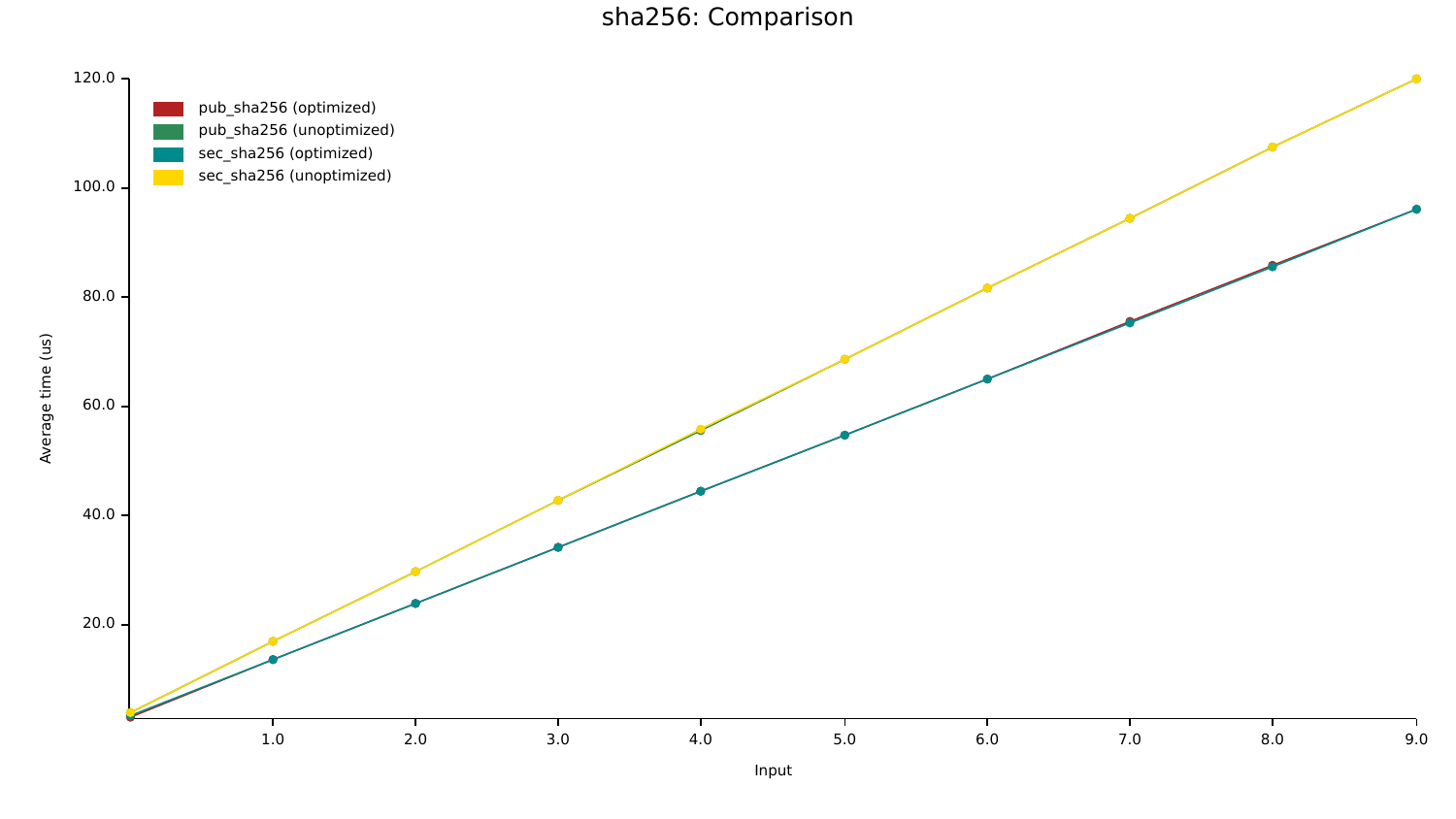}

\caption{A comparison between Cranelift running \texttt{ct-Wasm} and traditional Wasm versions of the same SHA-256 routine. The x-axis is the number of 256-byte chunks processed. Since we are able to take advantage of most performance optimizations other than peephole optimizations, the \texttt{ct-Wasm} version incurrs almost no overhead while granting constant-time guarantees.}
\label{fig:sha256perf}
  \end{mdframed}
\end{figure}

After running both traditional Wasm and \texttt{ct-wasm} versions of the same implementations of the SHA-256, Salsa20, and TEA crypptographic algorithms, we find that using \texttt{ct-wasm} leads to negligible performance impacts regardless of whether optimization is set to "None" or "Speed". (\autoref{fig:sha256perf}). We conjecture that the lack of overhead is because most optimizations apply equally to our DIT Cranelift IR and the original Cranelift IR, with the exception being peephole substitutions which happen to have negligible impact in this situation. 

\begin{figure}
\begin{mdframed}[linecolor=gray]
\centering
\includegraphics[width=\columnwidth]{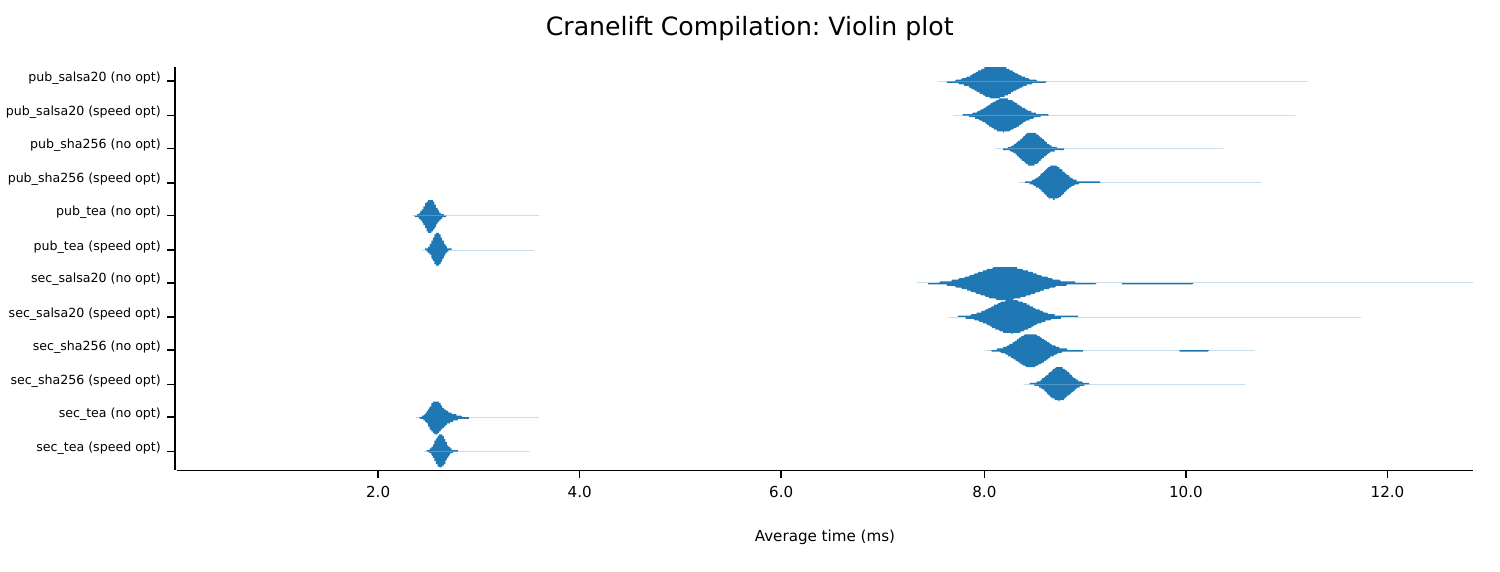}

\caption{A plot of compilation times between \texttt{ct-Wasm} and traditional Wasm variants of the same libraries. The compilation times for the traditional Wasm variants are shown on top. The compilation process also incurrs almost zero overhead after switching to \texttt{ct-Wasm}.}
\label{fig:compileperf}
\end{mdframed}
\end{figure}

Since Wasmtime is mostly used as a JIT compiler, the compilation time is also important. Regardless of which of the "None" or "Speed" optimization settings are used, we find that compilation overhead due to \texttt{ct-wasm} is negligible. (\autoref{fig:compileperf})

We ran the benchmarks on a Raspberry Pi 4, which uses four Cortex A72 out-of-order cores, running Ubuntu Server 20.04. Note that the Cortex A72 does not have support for \texttt{DIT}. (\autoref{sec:availability})

\subsubsection{Evaluating JANT}
\begin{figure}
\begin{mdframed}[linecolor=gray]

  \centering
\includegraphics[width=0.9\columnwidth]{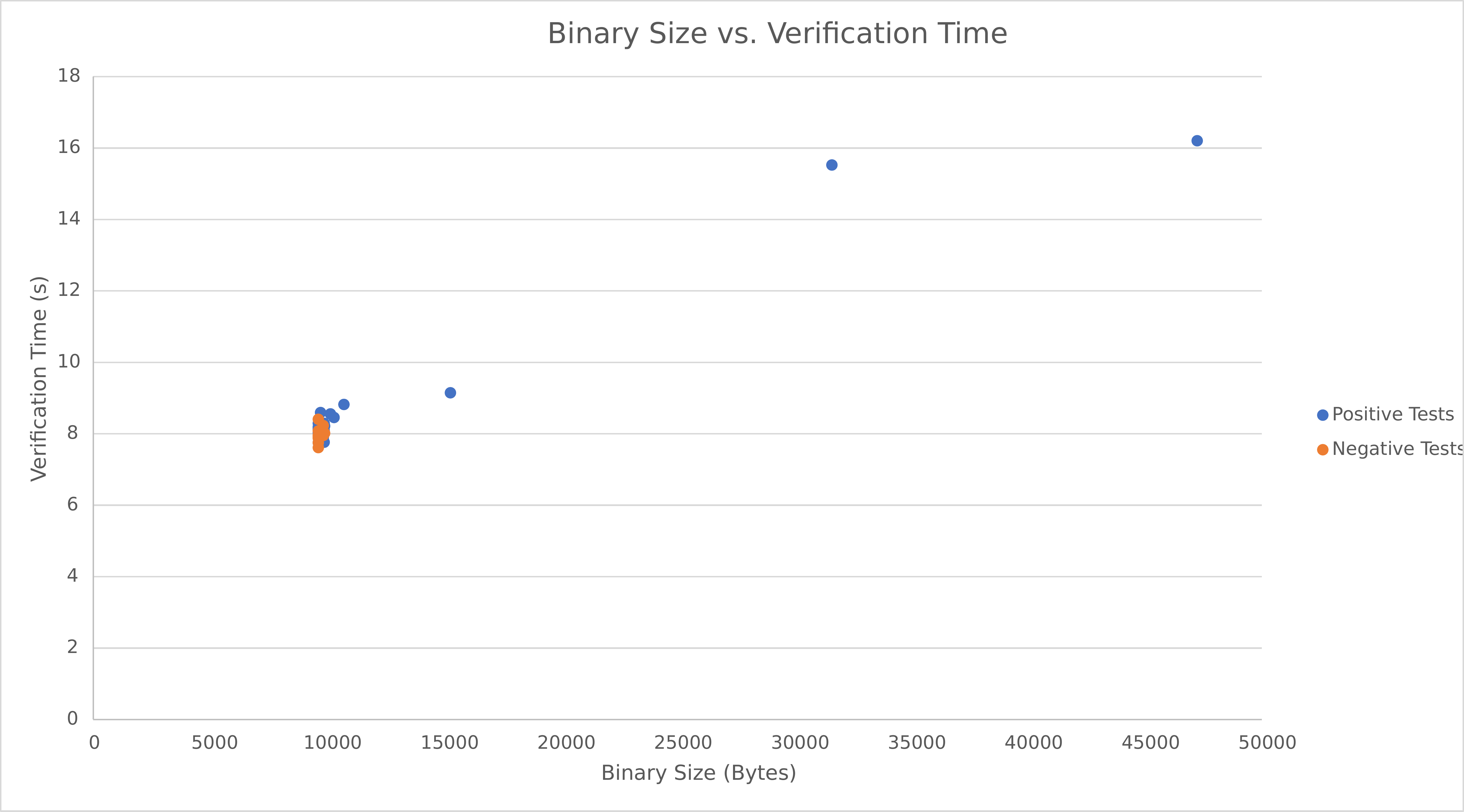}
\caption{A plot of binary size against verification time in JANT. The verification time includes all startup and teardown time, with Ghidra being run inside of a Docker container to facilitate automation. Much of JANT's runtime is dedicated to Ghidra's analysis routine, and the overall verification time seems to scale linearly relative to binary size.}
\label{fig:verifperf}
\end{mdframed}
\end{figure}

Our verifier does not need to be particularly fast, since in our intended use cases the verifier will rarely be run by the end user, mostly just Cranelift developers and constant-time library developers. However, we note that the verifier runs pretty quickly, with most "normal" programs verifying in just a few seconds. (\autoref{fig:verifperf})

We created a small test suite containing usable examples created by Watt et. al. \cite{watt2019ct} such as SHA-256, Salsa20, TEA, and a full port of TweetNaCl, in addition to simple positive and negative verification samples. Since a negative sample (that is, an \texttt{untrusted} function that violates constant-time semantics) is not expressible within \texttt{ct-wasm}, we perform some simple automatic binary patching to replace the JSON string being passed into the verifier. Overall, we tested the verifier against about 8,000 lines of \texttt{ct-wasm} code. 

We ran the benchmarks on a Ryzen 9 3900x system with 48 GB of RAM clocked at 3200MHz and an NVME SSD, on Ubuntu 20.04.1 LTS running in WSL2. 

\subsection{Takeaways from \texttt{ct-Wasmtime}}
With \texttt{ct-Wasmtime}, we are able to provide end-to-end verified constant-time execution for even fairly complex libraries such as full implementations of SHA-256, Salsa20, and a port of the TweetNaCl cryptographic library. A programmer can write their own cryptographic or otherwise timing-sensitive routines using \texttt{ct-wasm}, then generate and verify constant-time code that, barring speculative leaks, is guaranteed by the hardware to exhibit no timing leakages. 

However, there are a few drawbacks to programming directly in \texttt{ct-wasm}. 
\begin{enumerate}
\item WebAssembly's textual format is intended more as a compilation target than as a programming language; as a result, its style is very niche, with verbose and sometimes counterintuitive syntax. For example, a "loop" construct does not actually loop unless it concludes with a (br 0) instruction. In addition, WebAssembly text format is difficult to follow sometimes, with frequently occuring and lengthy keywords like \texttt{set\_global} and \texttt{get\_local} obscuring the logic. 
\item Writing \texttt{ct-wasm} requires the programmer to be constantly aware of which operations are constant-time and which are not, and write their code in a "constant-time" way; this presents a strange new paradigm for programmers new to constant-time programming, who may wonder why their code fails to compile when they add an if statement. 
\item Keeping secret and public memories separate is great for enforcing separation and preventing Spectre, but it becomes another burden on the programmer to keep track of which memory each pointer is pointing into, and more generally organize their own memory spaces in some way that maintains the separation of sensitivities. It also doesn't help that WebAssembly does not have any native array types, so allocating arrays in the linear memories is entirely left up to the developer. 
\end{enumerate}

We improve the development experience by adding a compiler backend for FaCT, an easy-to-learn C-like constant-time programming language, so programmers can write their code in FaCT and compile them to \texttt{ct-wasm} while keeping the constant-time semantics of their source program.

\section{FaCET}
FaCT (Flexible and Constant Time) is a C-like language with constant-time semantics, shown to be very easy to program in by Cauligi et. al. \cite{cauligi2017fact} Unfortunately, the constant-time semantics of FaCT extend only to the unoptimized LLVM IR, and the authors make no claims on the constant-timeness of the optimized IR or the machine code outputted by LLVM, so FaCT suffers from the same issue described in \autoref{sec:theproblem}.

In order to alleviate the issue, we present FaCET (Flexible and Constant-Time End-to-End Transpiler), an extension to FaCT that allows the programmer to target \texttt{ct-wasm} instead of LLVM IR. The result is that the user is able to write their timing-sensitive code in FaCT, then compile it through FaCET into \texttt{ct-wasm}, then feed the resulting output into \texttt{ct-Wasmtime} for verifiably constant-time execution on ARM. 

\subsection{FaCT Background}
FaCT looks very similar to C code, with native support for conditionals, looping, and arrays. \cite{cauligi2017fact} At a cursory glance, FaCT doesn't really look constant-time:
\pagebreak
\begin{verbatim}
export secret uint32 choose(
  secret bool cond,
  public uint32 a,
  public uint32 b
) {
  secret mut uint32 res = a;
  if(cond) {
    res = b;
  }
  return res;
}
\end{verbatim}

In the above example, note that there is clearly a secret-depdendent conditional branch on \texttt{cond}. However, the FaCT compiler will automatically perform transformations on the source program in order to generate constant-time LLVM bitcode. The above program will translate into a constant-time select, achieved with either XOR operations or inline assembly \texttt{cmov} instructions. \cite{cauligi2017fact}

The FaCT ecosystem suffers from the following issues:
\begin{enumerate}
  \item It makes assumptions that whichever assembly instructions LLVM decides to emit for its "constant-time" LLVM bitcode is actually constant time. For instance, if FaCT translates the above code into a special XOR-based select, it assumes that LLVM emits constant-time XOR and sign-extension operations. If FaCT translates the code to an inline assembly \texttt{cmov} instruction, it assumes both that \texttt{cmov} is constant-time on whichever chip the code ends up running on, and that LLVM does not generate non-constant-time code in order to set up and extract values to and from the inline assembly. 
  \item Its constant-time guarantees only extend down to the \textit{unoptimized} LLVM bitcode level. It is entirely possible for LLVM to optimize the code in a way such that it no longer exhibits constant-time behavior. \cite{cauligi2017fact}
\end{enumerate}

We solve both of these issues by rewriting the FaCT compiler's backend to output \texttt{ct-wasm} instead of LLVM bitcode, then feeding the outputted Wasm into \texttt{ct-Wasmtime} for compilation and execution.

\subsection{FaCT to \texttt{ct-wasm} Backend}
\label{section:factbackend}
Our compiler simply takes the place of the existing \texttt{codegen} module in FaCT's OCaml codebase. Since there is no equivalent \texttt{wasm} module for OCaml, we have to output the \texttt{wasm} manually. For ease of debugging, we choose to output \texttt{wast}, the "readable" WebAssembly text format, from the FaCET backend. Essentially, \texttt{wast} allows the stack-based WebAssembly to be expressed in infix form rather than postfix form. \cite{wat} (\autoref{fig:wat})

\begin{figure}
  \begin{mdframed}[linecolor=gray]
\begin{lstlisting}
(func $_crypto_stream_salsa20 untrusted (param $__v23_c i32) (param $__v75___v23_c_len i64) (param $__v24_n i32) (param $__v25_k i32) (result i32) (local $__rval i32) (local $__rctx s32) (local $__v26_kcopy i32) (local $__v27_input i32) (local $__v28_i i32) (local $__v88_lexpr i64) (local $__v89_lexpr i64) (local $__v29_ctimes i32) (local $__v30_j i32) (local $__v90_lexpr i64) (local $__v31_u s32) (local $__v32_i i32) (local $__v91_lexpr i64) (local $__v92_lexpr i64) (local $__v33_block i32) (local $__v34_remain i32) (local $__v93_lexpr i64) (local $__v94_lexpr i64) (local $__v35_cview i32) (local $__v36_i i32) (local $__v95_lexpr i64) (local $__v96_lexpr i64) (local $srbp i32) (local $prbp i32) (set_local $srbp (get_global $srsp)) (set_local $prbp (get_global $prsp)) (set_local $__rval (i32.const 0)) (set_local $__rctx (s32.const 1))(if (i64.eq (get_local $__v75___v23_c_len) (i64.const 0)) (then (set_global $srsp (get_local $srbp)) (set_global $prsp (get_local $prbp))  (return (i32.const 0))) (else ))(set_local $__v26_kcopy (set_global $srsp (i32.sub (get_global $srsp) (i32.mul (i32.const 8) (i32.div_u (i32.add (i32.const 7) (i32.mul (i32.const 1) (i32.wrap_i64 (i64.const 32)))) (i32.const 8)))))(call $memcpy_sec_sec (get_local $__v25_k) ...
\end{lstlisting}
\caption{An example of \texttt{wast} code generated by the FaCET backend. }
\label{fig:wat}
\end{mdframed}
\end{figure}

\subsubsection{Supported Features}
Our goal is to allow cryptographic and otherwise timing-sensitive algorithms like LibFTFP to be expressible in FaCT and translatable to \texttt{ct-wasm}. As such, we support 8-, 16-, 32-, and 64-bit signed and unsigned integers, conditionals, loops, arrays, and passing integers and arrays to functions. 

We decided not to support structs in our backend. The reason for this is because FaCT's structs were mainly included for interoperability with C code; however, \texttt{ct-wasm}'s separate secret and public memories make it difficult at runtime to craft structs containing both secret and public values in a way that \texttt{ct-wasm} can handle. We claim this added complexity negates the convenience benefits of having support for structs. 

We also decided not to support vector operations. This is because \texttt{ct-wasm} currently does not support the new Wasm SIMD proposal, so vector operations would not translate to SIMD operations. We consider it future work to extend \texttt{ct-wasm} to include support for the new SIMD extension of Wasm. 

\subsubsection{Handling Arrays}
WemAssembly does not allow for arrays to be allocated on the stack by design. Since all FaCT arrays are stack-allocated, we instead create a simulated stack on top of a linear memory; we initialize a simulated stack pointer (simply a global variable) to the top of the linear memory and add or subtract from this pointer when we need to create or destruct an array, respectively. 

Due to the separation of secret and public memories, we create two separate stacks -- a secret stack for holding secret ararys, and a public stack for holding public arrays. This allows us to inherit the Spectre resistance of \texttt{ct-wasm} even when the source program may contain seemingly Spectre-unsafe constructs like in \autoref{fig:spectrepht}.

\begin{figure}
\centering
\includegraphics[width=\columnwidth]{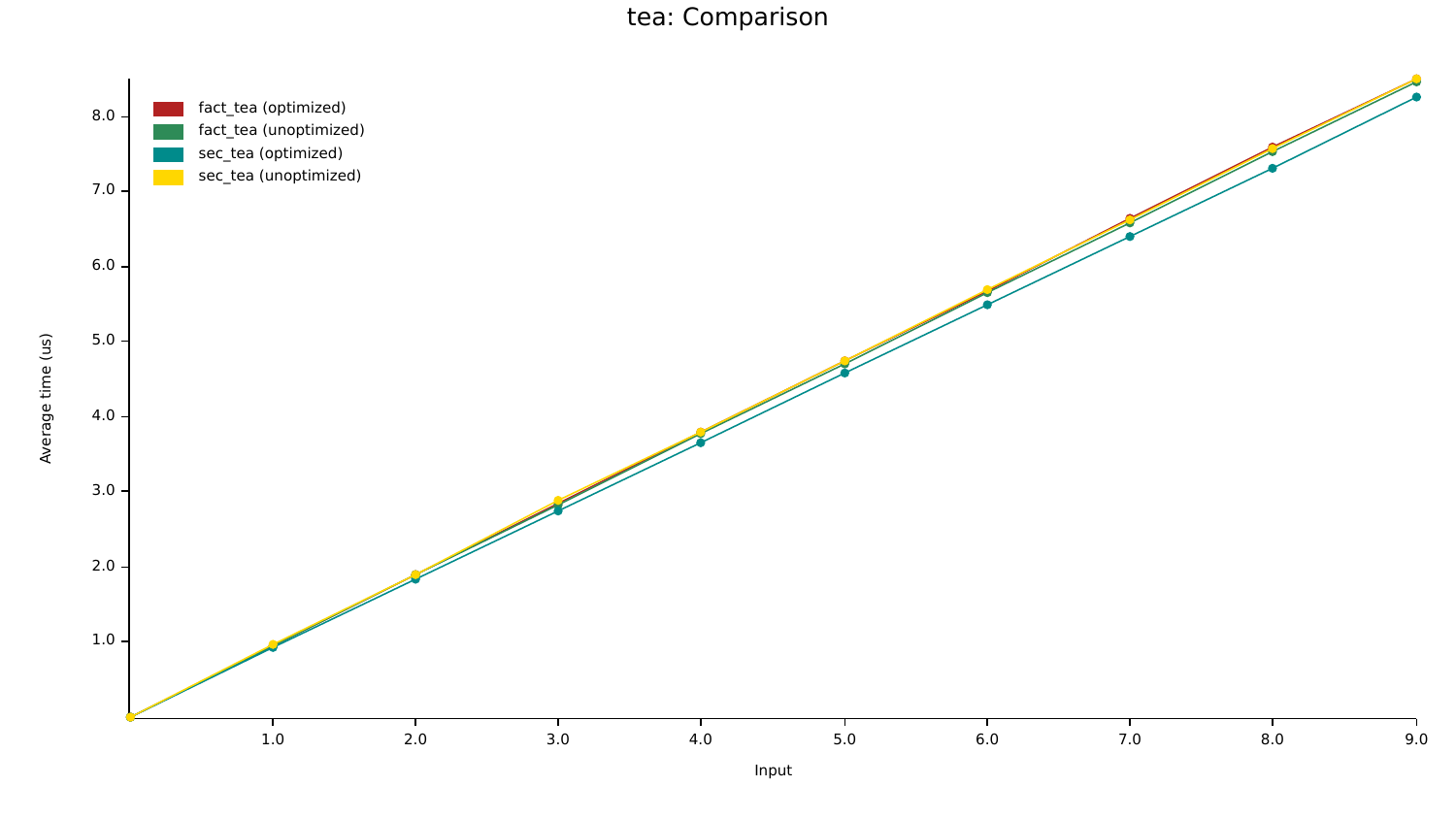}
\caption{A comparison between two versions of the same TEA routine, one implemented directly in \texttt{ct-wasm}, and the other translated to \texttt{ct-wasm} from FaCT. The x-axis is the number of blocks being encrypted and decrypted. We note that the performance overhead of FaCT is very small, reaching a maximum of 3.4\% when both the FaCT and \texttt{ct-wasm} versions are optimized. }
\label{fig:teafactperf}
\end{figure}

\begin{figure}
\centering
\includegraphics[width=\columnwidth]{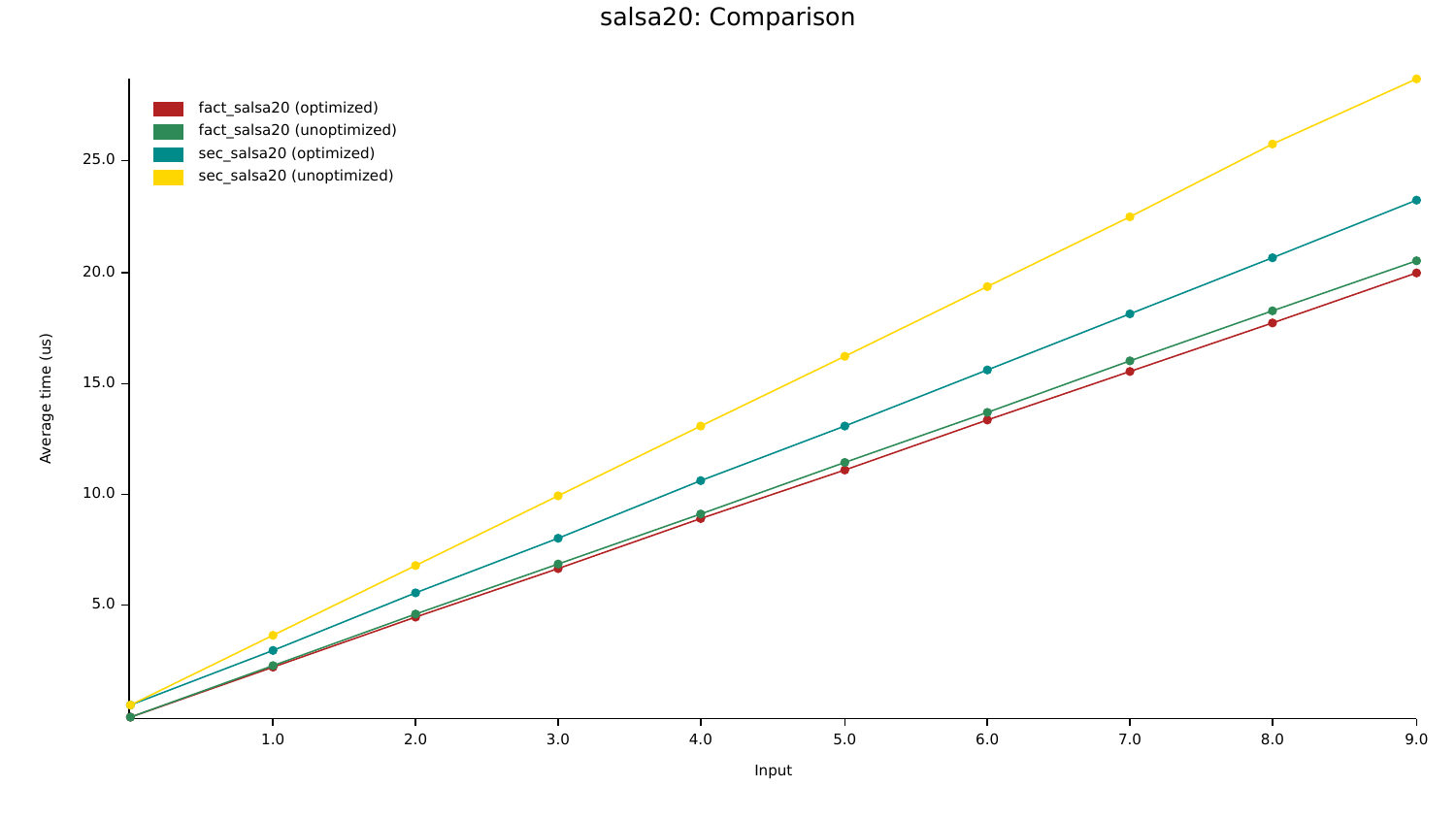}
\caption{A comparison between two implementations of the Salsa20 stream cipher, one implemented directly in \texttt{ct-wasm} by Watt, et. al. \cite{watt2019ct}, and the other implemented in FaCT by Cauligi et. al. \cite{cauligi2017fact} The x-axis is the number of 256-byte blocks being encrypted. }
\label{fig:salsa20factperf}
\end{figure}

\subsection{Evaluation of FaCET}
\subsubsection{How Do We Know FaCET Preserves Constant-Timeness?}
\label{sec:howfactct}
In \texttt{ct-wasm}, functions defined as \texttt{untrusted} \textit{must} be constant-time, since the type system ensures that unsafe operations (like calling \texttt{trusted} functions or invoking \texttt{declassify}) cannot be done by \texttt{untrusted} functions. Therefore, as a first step, we mark every function generated by FaCET as \texttt{untrusted}, with its arguments reflecting the secrecy traits of the input program. This encourages the user to look closer at their code and manually mark individual functions as trusted if the resulting program fails to type-check. As a result, we avoid issues such as unnoticed usages of \texttt{declassify} in the source program leading to the entire function losing its constant time guarantees. 

In addition, we make sure that the input program's secrecy traits are maintained. For example, \texttt{secret uint32} variables would translate to \texttt{s32} variables and \texttt{secret uint8[]} arrays would always be allocated in the secret memory. By simply examining function signatures, we can rest assured that the resulting \texttt{ct-wasm} will avoid secret-dependent timing by construction. 

\subsubsection{Performance}
Since the translation is fairly one-to-one, the performance of FaCET-generated \texttt{ct-wasm} should be similar to hand-written \texttt{ct-wasm}. After directly porting the \texttt{ct-wasm} implementation of TEA into FaCT and back into \texttt{ct-wasm}, we find that the FaCT version only reaches up to 3.4\% overhead when compared to the \texttt{ct-wasm} version. (\autoref{fig:teafactperf}). 

We also compared two implementations of the Salsa20 stream cipher, one written in \texttt{ct-wasm} by Watt, et. al. and the other written in FaCT by Cauligi, et. al. and extended slightly to match the other's interface. \cite{watt2019ct,cauligi2017fact} (\autoref{fig:salsa20factperf}) We note that counterintuitively, the FaCT version runs slightly faster in addition to being far easier to read and understand. We attribute the difference to the FaCT implementation being structured to pass values more effectively between functions and as a result completely avoiding several loads and stores from global memory addresses. 

\subsection{Related Work}
\textbf{Constant-Time Verification:} Verifying code for constant-time behavior is nothing new. \cite{almeida2016verifying, cauligi2020constant, barthe2018secure,bhargavan2013implementing,beringer2015verified, appel2015verification} As mentioned earlier, Almeida et al. have developed a software verifier (\texttt{ct-verif}) that verifies that a given piece of software is constant-time by checking for secret-dependent branching, loads/stores, and documented timing-variable operations (such as integer division). \cite{almeida2016verifying} In terms of Spectre-safe verification, Cauligi et. al. developed a symbolic execution tool, Pitchfork, that simulates speculative execution up to a specified window size and finds paths where secret-dependent data may be used in load/store addresses or branch targets, even speculatively. \cite{cauligi2020constant}

\textbf{Constructing Constant-Time Programs:} Our work is also related to other programs that attempt to compile or translate programs to make them constant-time. \cite{borrello2021constantine,braun2015robust,wang2017security} For example, the Constantine system automatically hardens programs against side channels using a dataflow linearization technique. \cite{borrello2021constantine} Braun et. al. used techniques such as time padding, isolating per-core resources, and lazy state cleaning to automatically mitigate Clang-compiled C/C++ libraries against side channels. \cite{braun2015robust} Barthe et. al. created a formally-verified C compiler that was proven to preserve constant-timeness throughout optimization passes. \cite{barthe2019formal}

\textbf{Hardware Verified Constant-Time:} Although we did not do any hardware-based verification, we do rely on verification done by ARM to ensure \texttt{DIT} guarantees actually hold. Gleissenthall et. al. showed an approach to eliminate timing side channels in hardware. \cite{gleissenthall2019iodine}. In addition, work is underway to add similar constant-time guarantees to the RISC-V specification. \cite{riscv} Finally, SecVerilog is a hardware design language to allow building systems with verifiable control over timing channels. \cite{zhang2015hardware}

\textbf{Safety of Wasm:} Johnson, et. al. also performed verification on Cranelift-generated Wasm, instead verifying SFI safety properties such as memory safety and isolation. \cite{johnson2021veriwasm} We analyzed our constant-time verification in the presence of speculative leaks. Swivel is another approach to harden Wasm against Spectre attacks using Intel's CET feature set. \cite{narayan2021swivel} 

\begin{figure*}
  \begin{mdframed}[linecolor=gray]

\begin{center}
\begin{tabular}{| c | c | c |}
\hline
Compilation Step & Ensuring Preservation\\
\hline\hline
FaCT & Semantics of FaCT \cite{cauligi2017fact}\\
\texttt{ct-wasm} & Semantics of \texttt{ct-wasm} \cite{watt2019ct} (\autoref{sec:howfactct})\\
Cranelift IR & Mapping to \texttt{DIT} IR instructions (\autoref{sec:craneliftchanges})\\
Optimized Cranelift IR & DIT-aware optimizations (\autoref{sec:compileropt})\\
AArch64 machine code & \texttt{DIT}-aware lowering, JANT verifier (\autoref{sec:verifying})\\
Code execution & \texttt{DIT} guarantees (\autoref{sec:ditdive})\\
\hline
\end{tabular}
\end{center}
\caption{A table listing the various steps we take to go from FaCT source code to machine code running on the processor. At each step, we consider how we can ensure the constant-time semantics of the original program are preserved.}
\label{fig:compilationprocess}
\end{mdframed}
\end{figure*}

\section{Conclusion and Future Work}
With FaCET, \texttt{ct-Wasmtime}, JANT, and \texttt{DIT}, we are able to soundly verify the preservation of constant time semantics through the entire process going from a high-level language, through several layers of compilation passes, down to the microarchitecture running raw machine code. (\autoref{fig:compilationprocess}). Our approach is relatively practical, opening the possibility for browsers and cloud providers to adopt a version of \texttt{ct-wasm} in the future using similar or identical techniques. 

We urge other CPU vendors to introduce some timing guarantees on their processors in a manner similar to ARM \texttt{DIT}, so that truly constant-time code can be run on them. We also urge compiler developers to document exactly when compiler optimizations can be trusted to preserve constant-time code, and offer solutions like secrecy annotations or compiler flags so that security-conscious software developers are not caught off guard. 

Finally, we hope to work on the following in the future:
\begin{enumerate}
  \item We would like to advance the constant-time WebAssembly proposal \cite{ctwasmspec} and pave the way for some version of \texttt{ct-wasm} to eventually make it into browsers and mainstream Wasm runtimes. There is significant demand for cryptography solutions within the Wasm sandbox. \cite{ctwasmdemand}
  \item We would like to add support for Wasm's new SIMD extensions in \texttt{ct-wasm}, and examine how this may affect support in \texttt{ct-Wasmtime}, JANT, and FaCET. Since we have not yet examined ARM's \texttt{DIT} guarantees on SIMD instructions, this would be a good opportunity to take a deeper look into them. 
  \item When ARM chips are released with the \texttt{PSTATE.DIT} feature, we would like to examine the runtime overhead of enabling the feature, whether from setting the bit itself or from executing affected instructions. 
  \item As part of \autoref{sec:compileropt}, we effectively removed peephole optimizations from being applied to \texttt{DIT} IR instructions. We would like to formally prove the constant-time-preserving properties of peephole optimizations and selectively allow verified peephole optimizations to operate on \texttt{DIT} instructions. 
\end{enumerate}

\bibliography{usenix.bib} {}
\bibliographystyle{plain}

\end{document}